\documentclass[a4paper,11pt]{article}
\usepackage{jcappub} % for details on the use of the package, please see the JINST-author-manual
\usepackage{lineno}
\usepackage{aas_macros}
\usepackage{gensymb}
\usepackage{booktabs}

\newcommand{\arcsec}{"}
\newcommand{\arcmin}{'}

\graphicspath{{./}{figs/}}

\title{Hard X-ray emission from blazars associated with high-energy neutrinos}

\author[a,b]{A. V. Plavin}
\author[c,d]{R. A. Burenin}
\author[e,b,f]{Y. Y. Kovalev}
\author[c]{A. A. Lutovinov}
\author[g,h,\dagger]{A. A. Starobinsky}
\note[$\dagger$]{Deceased}
\author[i,j]{S. V. Troitsky}
\author[c,d,i]{E. I. Zakharov}

\affiliation[a]{Black Hole Initiative at Harvard University, 20 Garden Street, Cambridge, MA 02138, USA}
\affiliation[b]{Astro Space Center of Lebedev Physical Institute, Profsouznaya str.~84/32, Moscow 117997, Russia}
\emailAdd{alexander@plav.in}
\affiliation[c]{Space Research Institute,
Profsouznaya str.~84/32, Moscow 119997, Russia}
\affiliation[d]{National Research University Higher School of Economics, Moscow 101000, Russia}
\affiliation[e]{Max-Planck-Institut f\"ur Radioastronomie, Auf dem H\"ugel 69, Bonn 53121, Germany}
\affiliation[f]{Moscow Institute of Physics and Technology, Institutsky per. 9, Dolgoprudny 141700, Russia}
\affiliation[g]{L. D. Landau Institute for Theoretical Physics RAS, Chernogolovka, Moscow region 142432, Russia}
\affiliation[h]{Bogolyubov Laboratory of Theoretical Physics, Joint Institute for Nuclear Research, Dubna 141980, Russia}
\affiliation[i]{Institute for Nuclear Research of the Russian Academy of Sciences, 60th October Anniversary Prospect 7a, Moscow 117312, Russia}
\affiliation[j]{Physics Department, Lomonosov Moscow State University, 1-2 Leninskie Gory, Moscow 119991, Russia}

\abstract{
Bright blazars were found to be prominent neutrino sources, and a number of IceCube events were associated with them. Evaluating high-energy photon emission of such blazars is crucial for better understanding of the processes and regions where neutrinos are produced. Here, we focus on hard X-ray emission observed by the SRG/ART-XC telescope, by the Swift/BAT imager, and by the INTEGRAL/IBIS telescope. Their energy range $\gtrsim10$~keV is well-suited for probing photons that potentially participate in neutrino production by interacting with ultrarelativistic protons.
We find that neutrino-associated blazars tend to demonstrate remarkably strong X-ray emission compared to other VLBI blazars in the sky.
Both neutrinos and hard X-rays are found to come from blazars at cosmological distances $z\sim1$, and are boosted by relativistic beaming that makes it possible to detect them on Earth. Our results suggest that neutrinos are produced within compact blazar jets, with target X-ray photons emitted from accelerated jet regions.
}

\begin{document}

\maketitle
\flushbottom

\section{Introduction}

While high-energy astrophysical neutrinos above TeV have been reliably detected by IceCube \cite{IceCube-diffuse-discovery,IceCube-diffuse-recent}, ANTARES \cite{ANTARES-diffuse} and Baikal-GVD \cite{Baikal:diffuse}, their origin remains a subject of debates \cite[see, e.g.,][]{Meszaros-2017review,Troitsky-UFN2021}. However, in recent years, considerable progress has been made in determining the sources of high-energy neutrinos.

Indeed, similarly to the electromagnetic radiation, neutrino emission may be boosted if it is produced in relativistic jets pointing towards the observer. Active galactic nuclei (AGN) with such jets are called ``blazars'', and were suspected to produce astrophysical neutrinos long before the latter were discovered. The association of neutrino events with one of the blazars, TXS~0506$+$056, see \cite{IceCube-0506HE}, marked the start of observational confirmation of those expectations. In this work, we use the term ``blazar'' to refer to AGNs with jets pointed towards us at a viewing angle of several degrees, following the unified scheme in \cite{1995PASP..107..803U}. This naming is not directly based on any specific spectral, optical, gamma-ray or other properties.

\begin{figure*}
\centering
\includegraphics[width=0.85\linewidth,trim=0mm -3mm 0mm 0mm]{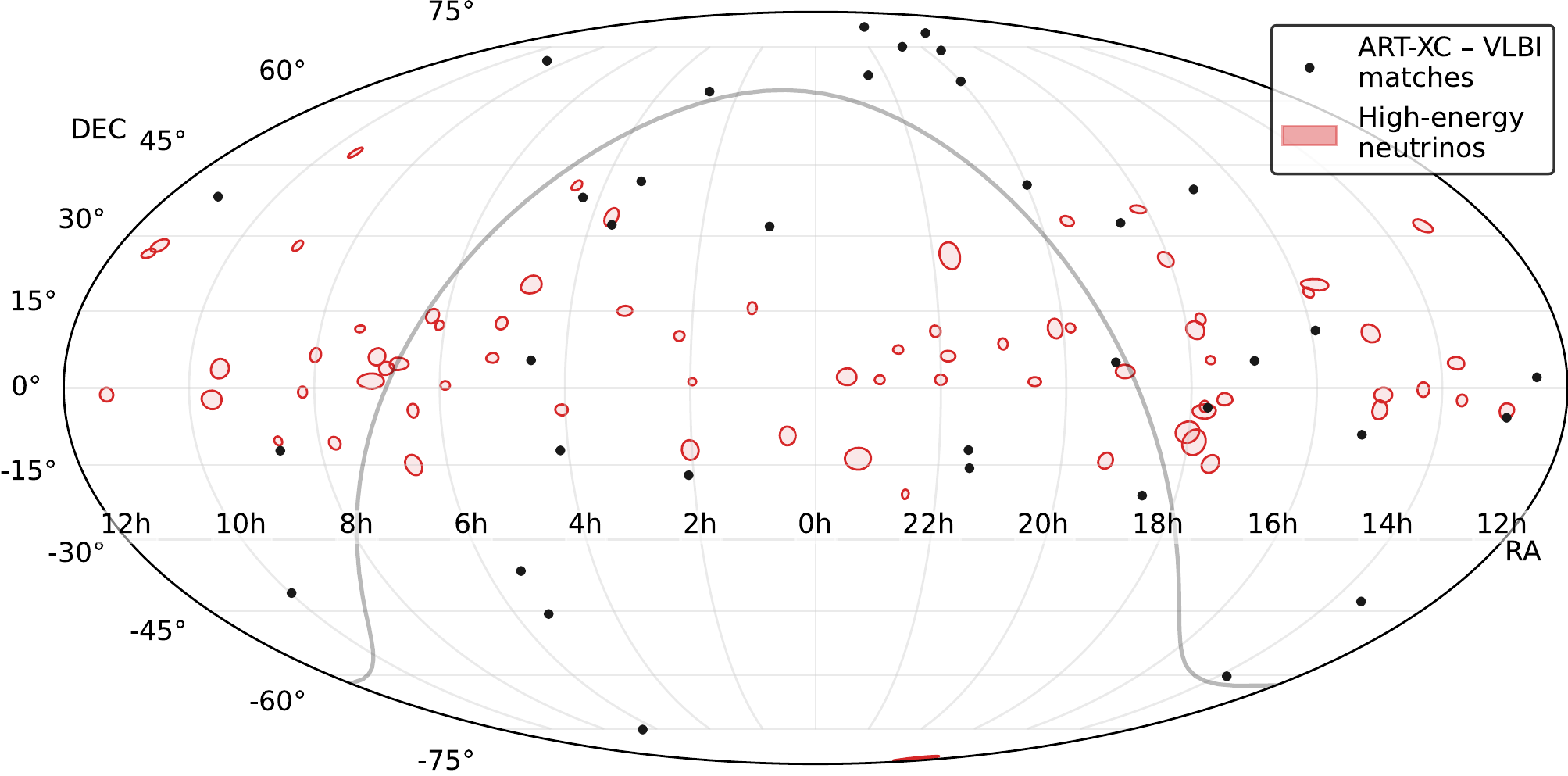}
\includegraphics[width=0.85\linewidth,trim=0mm -3mm 0mm 0mm]{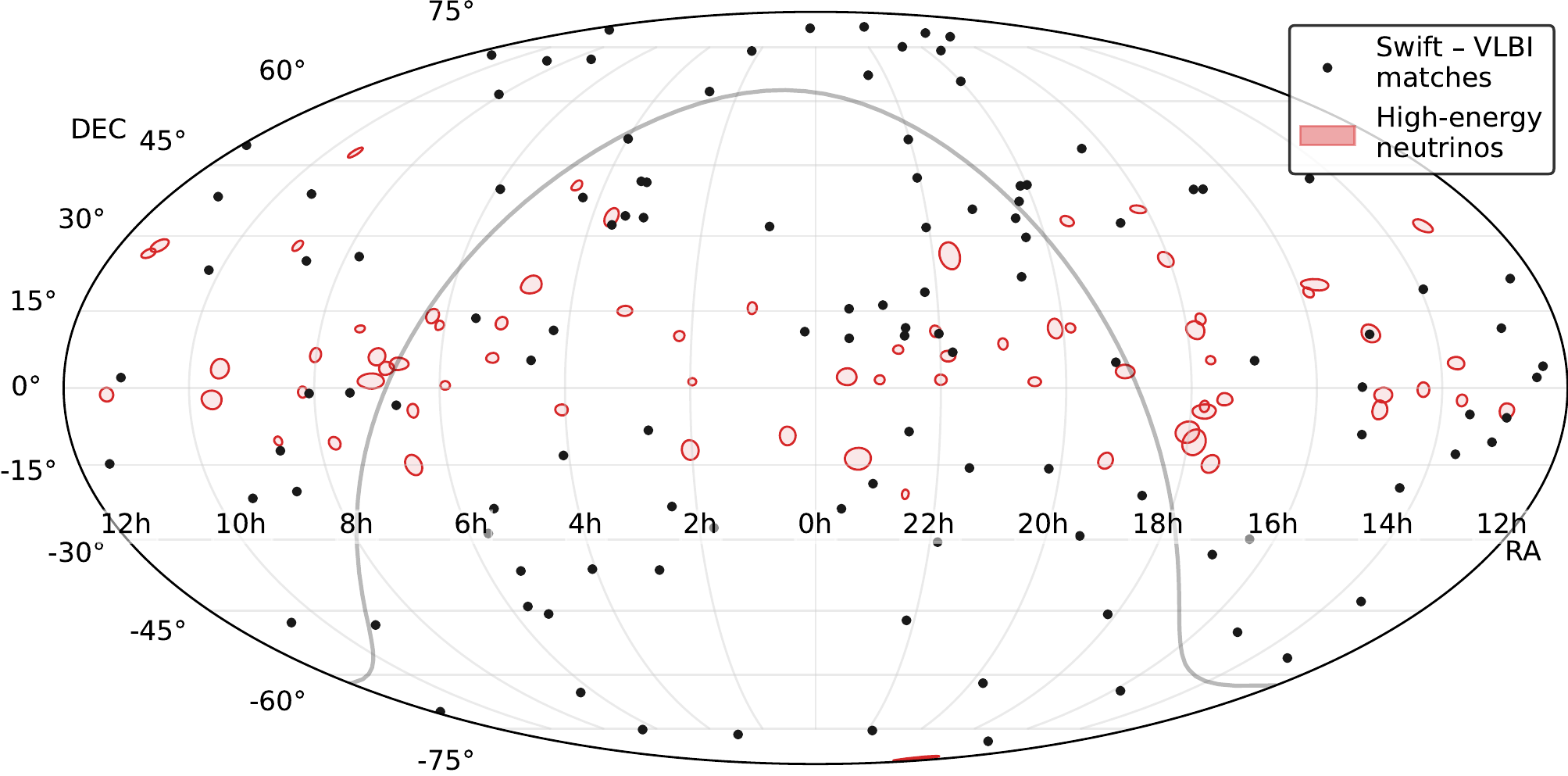}
\includegraphics[width=0.85\linewidth,trim=0mm -2mm 0mm 0mm]{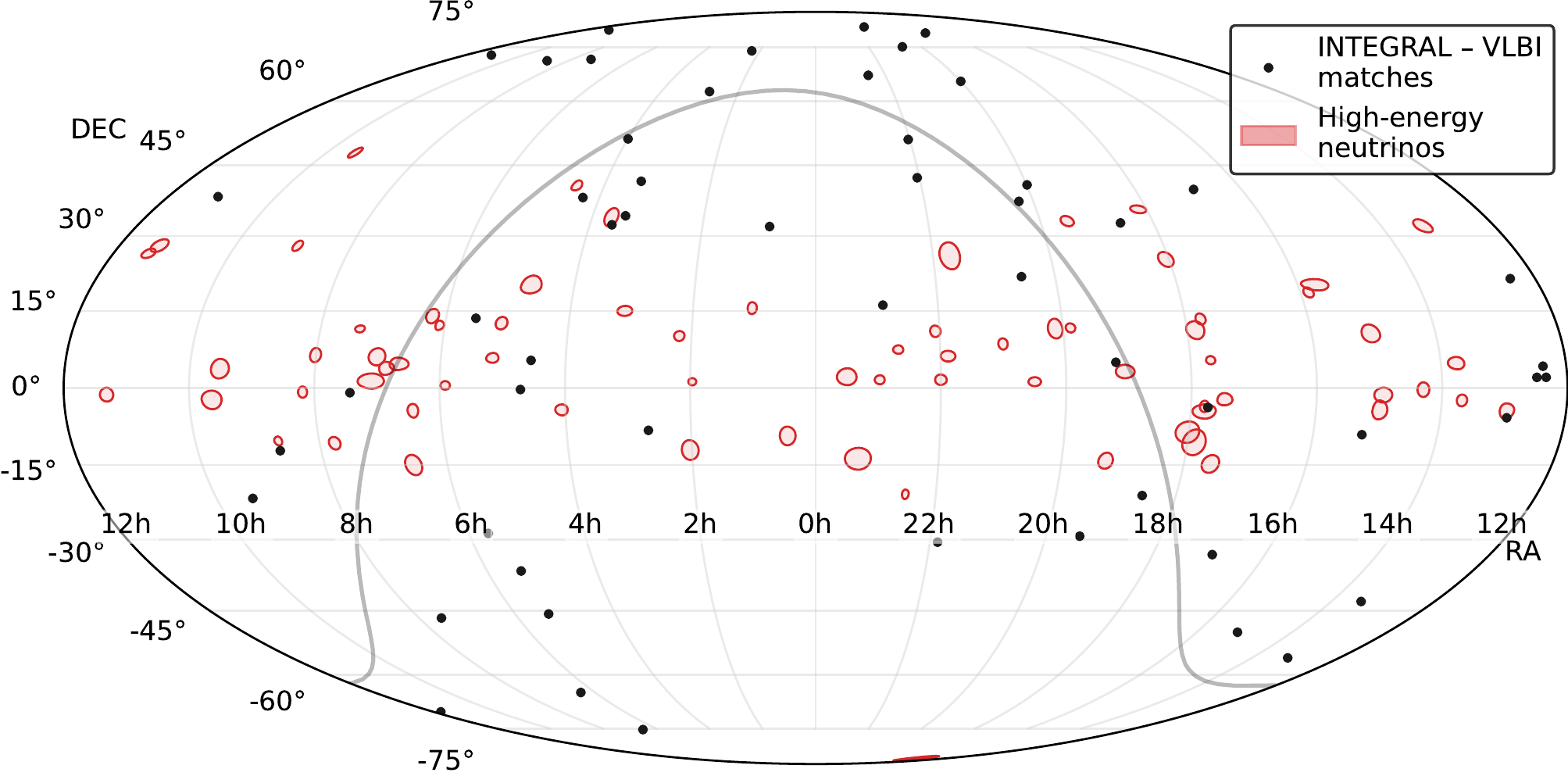}
\caption{
\label{f:skymap}
Sky distributions of X-ray sources (black dots) matched with VLBI blazars, and of IceCube neutrino detections. Neutrino error regions are shown as red ellipses following \autoref{s:data}. See \autoref{s:data} and \autoref{t:datasets} for source selection criteria.
}
\end{figure*}

One of the most direct observational indications of a Doppler-boosted relativistic parsec-scale jet in an AGN is the presence of a strong, extremely compact radio source. These sources can be selected with Very-Long-Baseline Interferometry (VLBI) thanks to its high angular resolution. We further motivate using VLBI-selected samples as blazar catalogs and ensure their uniformity in \autoref{s:Xblazars}.
Remarkably, the observed neutrino associations with blazars exhibiting bright, compact radio emission have grown significantly in recent years. \cite{neutradio1} found a significant excess of bright VLBI-selected radio blazars close to the arrival directions of published IceCube events with energies above 200~TeV. The significance of the excess continues to grow with increasing numbers of detected neutrinos \cite{neutradio2023}. Support for the neutrino association with radio blazars is starting to come from ANTARES \cite{2023arXiv230906874A} and Baikal-GVD \cite{Baikal-0506} experiments. Additionally, \cite{neutradio1} found that these high-energy neutrino events coincide with radio flares of the associated blazars, for which radio light curves are available. This result was confirmed by \cite{Hovatta-2021} with different radio data. Finally, \cite{neutradio2,2024MNRAS.527L..26S} found that the same sources --- blazars --- are significantly correlated with lower-energy neutrinos as well, thus extending the association to lower energies of a few TeV. The combined statistical significance of analyses below and above 200~TeV, for which independent data were used, reaches 4.3 standard deviations \cite{neutradio2023}.
Neutrino-blazar associations or hints thereof were also reported by \cite{2016NatPh..12..807K,Elisa:blazars,Anna:blazars,Sara:blazars22,Sara:blazars23}. Some other studies do not find a significant correlation \cite[see, e.g.,][]{Zhou,2023ApJ...954...75A}. They do not necessarily contradict the presence of associations, though; see discussion in \cite{Plavin-on-Zhou}. The neutrino sky is likely diverse and includes a galactic contribution as well \cite{2022ApJ...940L..41K,2023PhLB..84137951A,2023Sci...380.1338I,2023MNRAS.526..942A}, but blazars appear to be the most prominent point-like neutrino sources.

The reported associations point to the central parsecs of blazars as the sites of high-energy neutrino production. However, a detailed understanding of the process is still out of reach, and even a particular localization of the emission region within these central parsecs has yet to be determined \cite[see, e.g.,][]{Neronov:RadioModel,neutradio2,Polina:RadioModel,2018ApJ...865..124M,2024arXiv240406867K}.

Neutrinos with energies above GeV are expected to be produced in interactions of accelerated protons $p$ with ambient photons $\gamma$ or matter \cite[e.g.,][and references therein]{CaponeLipariVissaniReview,Troitsky-UFN2021}. Central parsecs of AGNs are rich in radiation but relatively poor in baryonic matter, so, quantitatively, the probability of a $p\gamma$ interaction normally exceeds that of a $pp$ one (e.g., \cite{Boettcher-rev}; see, however, \cite{Neronov:RadioModel}). The cross section of the $p\gamma$ interaction peaks at the energies which allow for the resonant production of $\Delta^+$ baryon. Neutrinos are born in decays of $\pi^+$ mesons present among the reaction products; $\pi^0$ mesons decay to energetic secondary photons. Simple relativistic kinematics \cite[e.g.][]{CaponeLipariVissaniReview}, suggests the following benchmark relations between the energies of the incoming proton, $E'_{\rm p}$; of the target photon, $E'_\gamma$; and of the produced neutrino, $E'_\nu$,
\begin{equation}
E'_{\rm p} \simeq 20 E'_\nu; ~~~~ E'_{\rm p} E'_\gamma \simeq m_\Delta^2\,,
    \label{benchmark_primed_energies}
\end{equation}
where $m_\Delta\approx 1.23$~GeV is the $\Delta^+$ mass in particle-physics units. Hereafter, the primes mean the energies are in the source frame, where target photons are assumed to be isotropic. These relations work for order-of-magnitude estimates in most cases, though should be used with caution in detailed studies \cite{VissaniNot20Enu}.

Relations~(\ref{benchmark_primed_energies}) suggest that the target photons required to produce $\sim 10 - 1000$~TeV neutrinos should have energies in the X-ray band (see \autoref{s:disc:mech} and \autoref{f:spectrum} for more details). These X-rays can escape the AGN central parsecs freely and therefore be observed. As we discuss below, the observed X-ray fluxes should be higher if the emission is Doppler boosted. Thus, target photons are more likely to be detected if they are produced in the parsec-scale jet of a blazar, compared to the accretion disk corona or other regions in the immediate environment of the central black hole.

The purpose of the present work is to provide additional observational details to the multimessenger picture, important for choosing a specific model. We search for observational indications to these high hard X-ray fluxes of neutrino-associated blazars. 
The rest of the paper is organized as follows. First, in \autoref{s:data}, we define and describe the utilized neutrino, X-ray, and VLBI observational datasets and their characteristics. Then, \autoref{s:statall} walks through the statistical analysis performed to evaluate the correlation between neutrinos and X-ray blazars and presents its results. In \autoref{s:disc}, we put our observational findings into a broader multimessenger context, discussing the mechanisms responsible for neutrino production and those that aid in actually detecting these neutrinos. Finally, in \autoref{s:summary}, we summarize our results and outline the likely near-future perspectives for the neutrino associations.

\section{Data}
\label{s:data}

\begin{figure}
\centering
\includegraphics[width=0.65\linewidth]{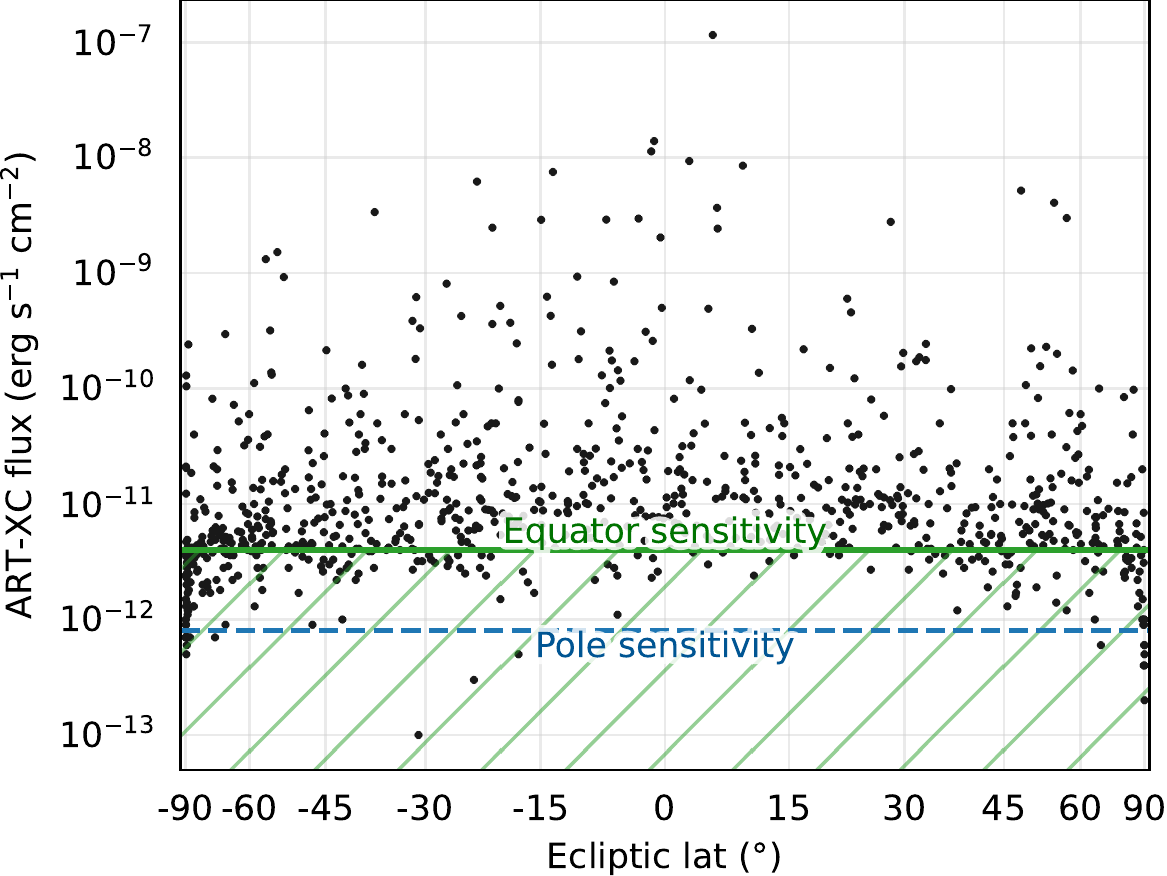}
\caption{
The sensitivity limits of the X-ray survey by the SRG/ART-XC telescope, for different regions by the Ecliptic latitude. Values for the Ecliptic pole and equator regions are shown with horizontal lines. Black dots represent measured fluxes of individual sources from the catalog. We only use sources above the equatorial sensitivity limit for our analysis to make the sample as uniformly complete as possible, see \autoref{s:data}.
\label{f:artxc_sensitivity}
}
\end{figure}

\begin{table*}
\caption{
Summary of all datasets we utilize in our analysis. See \autoref{s:data} for a detailed description of each catalog and the motivation to choose these selection cutoffs.
}
\label{t:datasets}
\centering
\scriptsize
\begin{tabular}{lcccccc}
\hline\hline
Observatory & Energy or & \multicolumn{3}{c}{Number of catalog entries} & \multicolumn{2}{c}{Selection cutoffs}\\
            &       frequency              & Total  & $\geq$~min.~flux & VLBI matches      & Min. flux                                   & Max. separation\\
(1)         & (2)                 & (3)    & (4)              & (5)               & (6)                                            & (7)\\
\hline
IceCube\footnote{XXX}    & $\geq 200$~TeV      & 71     & --               & --                & --                                             & --\\
\hline
VLBI        & 8~GHz               & 21906  & 3243             & --                & 150~mJy                                        & --\\
\hline
   &             &      &                &                  & (erg\,s$^{-1}$\,cm$^{-2}$)   & \\
SRG/ART-XC  & 4..12~keV           & 867    & 663              & 36                & $4 \cdot 10^{-12}$   & 30$''$\\
Swift/BAT   & 14..195~keV         & 1632   & 1389             & 116               & $8.4 \cdot 10^{-12}$ & 10$'$\\
INTEGRAL/IBIS & 17..60~keV          & 929    & 746              & 55                & $1.4 \cdot 10^{-11}$ &  5$'$\\
ROSAT       & 0.1..2.4~keV        & 135118 & 71408            & 521               & $3\cdot 10^{-13}$    & 20$''$\\
\hline
\end{tabular}

\footnotesize
\flushleft
\textsuperscript{1} The IceCube high-energy event sample utilized for the analysis in this paper covers the period from 2009 to 2022, see details in \autoref{s:data_icecube}. Comparison with the recent IceCat-1 catalog that is based on different event reconstructions is given in \autoref{s:icecat}.

\end{table*}

\subsection{IceCube events}
\label{s:data_icecube}

We follow the earlier works studying blazars as neutrino sources \cite[most recently,][]{neutradio2023,2024MNRAS.527L..26S}, and utilize the same selection criteria for the highest-energy IceCube events. All alerts and alert-like events with the reported 90\% error region area below $10$~sq.~deg., and either belonging to the EHE category or having the best-fit estimated neutrino energy $E\geq200$~TeV are included. The resulting set of 71 tracks is collected in \cite{neutradio2023}, and includes events from the start of IceCube observations in 2009 to November 1, 2022. Directions of IceCube neutrinos in the sky are shown in \autoref{f:skymap}, and the summary of this sample is given in \autoref{t:datasets} that lists all datasets utilized in this paper.

Following comparison with radio blazars in \cite{neutradio1,neutradio2023}, we extend the IceCube reported uncertainties to account for systematic errors not included in the original estimates. According to \cite{2023arXiv230401174A}, reported uncertainties in RA and Dec correspond to full two-dimensional coverage. In this interpretation, the optimal approach is simply adding $0.78\degree$ \cite{neutradio2023}. Note that in earlier works we performed a two-step calculation: multiply errors by 1.3 and add $0.45\degree$ afterward. All results presented in this paper remain the same under this approach, both the statistical significance levels and the list of probable associations.

\subsection{X-ray sky surveys}

In this work, we focus on high-energy electromagnetic emission. Details for survey catalogs are summarized in \autoref{t:datasets}.

A relevant all-sky survey with the maximum sensitivity at around $\sim10$~keV was recently conducted using the Mikhail Pavlinsky ART-XC telescope \cite{2021A&A...650A..42P} onboard the SRG space observatory \cite{2021A&A...656A.132S}. Here, we use the one-year SRG/ART-XC all-sky survey catalog, which comprises 867 sources detected on the combined map of the first two 6-month scans of the sky \cite{2022A&A...661A..38P}. 

At higher energies, we use the survey made with the Burst Alert Telescope (BAT) on the Neil Gehrels Swift Observatory \cite[hereafter Swift/BAT,][]{2005SSRv..120..143B}. We use the latest 105-month Swift/BAT all-sky hard X-ray survey catalog \cite{2018ApJS..235....4O}, which contains 1632 hard X-ray sources. For our study, we additionally use data from the 17-yr hard X-ray all-sky survey catalog obtained by the IBIS (Imager on Board the INTEGRAL Satellite, \cite{2003A&A...411L.131U}) telescope onboard the INTEGRAL observatory, which includes a total of 929 hard X-ray sources \cite{2022MNRAS.510.4796K}.

In soft X-rays, we rely on the ROSAT all-sky survey data \cite[RASS,][]{1999A&A...349..389V}. We use the latest and most complete version of the RASS source catalog, which consists of more than $10^5$ sources detected in the 0.1--2.4~keV band \cite[2RXS,][]{2016A&A...588A.103B}.

For statistical analysis, it is important that the source catalog is as uniformly complete across the sky as possible. However, sensitivity varies significantly with the direction in all the surveys listed above. To eliminate non-uniformity to first order, we keep only sources with fluxes above a certain threshold. This threshold is chosen as the source detection limit in the most underexposed regions of the sky.

Because of similar sky survey strategies, the highest sensitivity for both RASS and ART-XC all-sky surveys is achieved at the Ecliptic poles and the lowest one is at the Ecliptic equator. To achieve approximate uniformity, we keep only sources with fluxes above the equatorial sensitivity limits, as illustrated in \autoref{f:artxc_sensitivity} for the SRG/ART-XC all-sky survey. The sensitivity patterns of the Swift/BAT and especially INTEGRAL/IBIS sky surveys are more complicated, see references above. We set the source selection threshold for these surveys conservatively so that the real sensitivity is better than the threshold in more than 90\% of the sky. Still, noticeable nonuniformity remains in the INTEGRAL/IBIS catalog (\autoref{f:skymap}); this is discussed in the context of our analysis in \autoref{s:stat}.
The minimum flux thresholds used in our study for each X-ray catalog are given in \autoref{t:datasets}.

\subsection{VLBI-selected radio blazars}

To uniformly distinguish blazars from all celestial sources, we rely on VLBI observations (\autoref{s:Xblazars}). A complete sample of VLBI-selected blazars, along with their positions and parsec-scale flux densities, is compiled in the Radio Fundamental Catalogue\footnote{\url{http://astrogeo.smce.nasa.gov/sol/rfc/rfc_2023a/}} (RFC). VLBI observations were performed at the 8~GHz band, including geodetic VLBI \cite{2009JGeod..83..859P,2012A&A...544A..34P,2012ApJ...758...84P}, the Very Long Baseline Array (VLBA) calibrator surveys (VCS; \cite{2002ApJS..141...13B,2003AJ....126.2562F,2005AJ....129.1163P,2006AJ....131.1872P,2007AJ....133.1236K,2008AJ....136..580P,r:wfcs,2016AJ....151..154G}), and other 8~GHz global VLBI, VLBA, EVN (the European VLBI Network), and LBA (the Australian Long Baseline Array) observations 
\cite{2011AJ....142...35P,2011AJ....142..105P,2011MNRAS.414.2528P,2012MNRAS.419.1097P,2013AJ....146....5P,2015ApJS..217....4S,2017ApJS..230...13S,2019MNRAS.485...88P,2021AJ....161...88P}.
The complete flux density-limited sample of VLBI-selected blazars consists of 3412 objects with a historic median 8-GHz flux density $S^\mathrm{VLBI}_\mathrm{8\,GHz}>150$~mJy integrated over their VLBI images. This blazar sample is used in the analysis throughout the paper.
While RFC collects VLBI flux density values at many radio frequencies, we chose 8~GHz due to the superior completeness characteristics of the sample at this band.

Available blazar redshifts were gathered from the NASA Extragalactic Database (NED).

\begin{figure}
\centering
\includegraphics[width=0.6\columnwidth,trim=0mm 1mm 0mm -1mm]{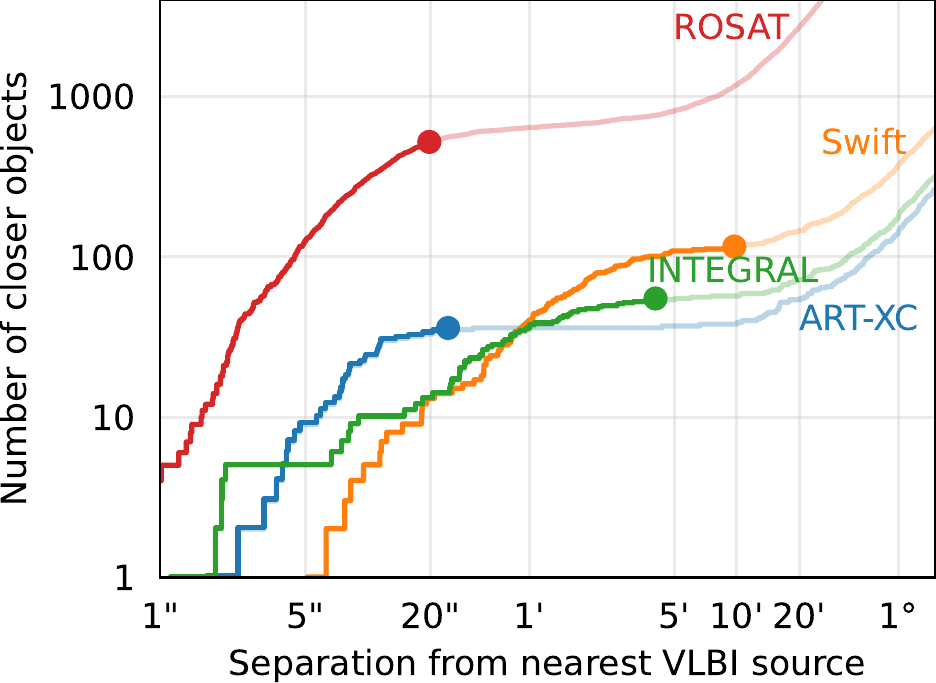}
\caption{
Angular separations between sources from X-ray catalogs and radio blazars observed by VLBI. VLBI positions are accurate at milliarcsecond scales, and this plot is interpreted as a demonstration of X-ray surveys positional accuracy.
The maximum separations to consider sources as matching are defined in \autoref{s:Xblazars} and \autoref{t:datasets}, and are marked as circles in this plot.
\label{f:separations}
}
\end{figure}

\section{Statistical analysis}
\label{s:statall}

\subsection{Selecting blazars from X-ray surveys using VLBI}
\label{s:Xblazars}

The first step of our analysis is to select the blazars from the X-ray catalogs. An observationally direct and uniform way to do this is by cross-matching X-ray sources with the complete VLBI sample (\autoref{s:data}). VLBI-bright sources are predominantly blazars: they exhibit strongly beamed jet emission \cite{2003ApJ...599..105L,2019ApJ...874...43L}, with a median Doppler factor close to 10 within the MOJAVE sample \cite{2021ApJ...923...67H}. Further, radio interferometry allows for a uniform sky coverage, unlike some other blazar selections \cite[e.g., the widely used, pure but non-uniform BZCAT catalog,][]{2015Ap&SS.357...75M}. A special multi-decade effort was dedicated to observing uniform complete samples; see \autoref{s:data} for details on the VLBI observational programs.

We assign an X-ray source to a VLBI blazar if their angular separation is smaller than a certain threshold. The threshold depends on the resolution of the X-ray telescopes and affects the completeness and purity of the blazar selection. To illustrate this, we show the distribution of X-ray~--~VLBI angular separations in \autoref{f:separations}. The number of chance coincidences is very low at small angular separations, most of them should be real associations. This distribution agrees very well with the angular resolution of the corresponding telescopes. The X-ray source position measurement accuracy should not be larger than 20\arcsec\ for ROSAT, 30\arcsec\ for SRG/ART-XC, 5\arcmin\ for INTEGRAL/IBIS,
and 10\arcmin\ for Swift/BAT.
With these thresholds, we expect to find no chance X-ray source~--~blazar coincidences for the ART-XC catalog, $\approx1$ chance coincidence for both INTEGRAL and ROSAT samples, and $\approx 10$ for Swift.

Following this procedure, we obtain 36 X-ray~--~VLBI matches in the SRG/ART-XC catalog. The matches for all surveys are visualized in \autoref{f:skymap}, and their counts are collected in \autoref{t:datasets}. We observe that the fraction of nearby matches with redshifts $z\lesssim0.2$ is higher than for objects in the complete VLBI sample: see \autoref{f:redshifts} in \autoref{s:disc:evolution}. Many of these objects still exhibit beamed jet emission, as indicated by that figure, but the beaming is weaker than in typical distant blazars. These objects are more likely to belong to other AGN classes, such as Seyferts and radio galaxies. Still, for our main statistical analysis, we treat all matches the same regardless of Doppler factors or distances to ensure uniformity. We further discuss the implications of the distance distribution in \autoref{s:disc:evolution}.

\subsection{Neutrino~--~X-ray emission correlation}
\label{s:stat}

Our aim is to investigate potential associations between neutrinos and blazars exhibiting strong hard X-ray emission. We determine the spatial correlation by counting pairs of neutrino~--~blazar matches within the neutrino localization errors as defined in \autoref{s:data}. This is a commonly used approach based on the correlation function, recently applied for neutrino-blazar correlation in \cite{2023arXiv230906874A}. Correlation significance is evaluated through Monte-Carlo simulations: we generate random $10^5$ realizations by scrambling the Right Ascension of all IceCube events while keeping their Declinations constant. This procedure is motivated by the IceCube location on the South Pole: its sensitivity only depends on Declination \cite[see, e.g.,][]{IceCube:7yr-sources,IceCube-0506HE,neutradio1}. As a result, these realizations reproduce the distribution of neutrino events, assuming that their arrival is not related to blazars in any way. Further, we count neutrino-blazar matches in each random realization in the same way as in the real data, and compute the $p$-value as the fraction of realizations with at least as many matches as in the real data.

For the main correlation analysis, we use the 36 X-ray~--~VLBI matches from the SRG/ART-XC catalog (\autoref{s:data}) as our blazar sample. We focus on the ART-XC survey due to the all-sky survey nature of its observations that ensures the uniformity of the catalog. After evaluating the directional correlation strength for ART-XC matches, we perform the same computations for X-ray~--~VLBI sources selected from other X-ray surveys for comparison.

\begin{table*}
\caption{
Summary of our statistical analyses addressing associations between IceCube neutrinos and bright X-ray blazars.
}
\label{t:pvalues}
\centering
\scriptsize
\begin{tabular}{lrrrrrp{17em}}
\hline\hline
Observatory & \multicolumn{2}{c}{Statistical significance} & \multicolumn{3}{c}{\# of sources} & List of neutrino matches\\
\cmidrule(lr){2-3}\cmidrule(lr){4-6}
 & p$-$value & Equivalent & Total & \multicolumn{2}{c}{Neutrino matches}\\
 & (\%) & ($\sigma$) & & Actual & By chance & \\
(1) & (2) & (3) & (4) & (5) & (6) & (7) \\
\hline
   SRG/ART-XC & 0.5 & 2.8 &   36 &  4 &  0.7 & PKS~1741$-$038$\times2$, 3C~279, NRAO~140\\
Swift/BAT & 4 & 2.0 &  116 &  5 &  1.9 & 3C~279, NRAO~140, TXS 1502+106, TXS~2157+102, TXS~2145+067\\
 INTEGRAL/IBIS & 1.0 & 2.6 &   55 &  4 &  0.7 & PKS~1741$-$038$\times2$, 3C~279, NRAO~140 \\
    ROSAT & 12 & 1.6 &  521 & 12 &  8.2 & PKS~1741$-$038$\times2$, 3C~279, NRAO~140, TXS~2145+067, 3C~57, J1733$-$1304, J1310+3220, J0505+0459, J2254+0054, J0509+0541, J1451$-$0127\\
\hline
\end{tabular}
\footnotesize
\flushleft
Notes:
\item For each X-ray survey catalogue intersected with the VLBI catalogue, we list matches with IceCube neutrinos in columns (4), (5), (6), and statistical significance estimates in columns (2) and (3). The p-value is the probability to observe this many matches by chance alone.
\item Notable neutrino coincidences are visualized in \autoref{f:skymaps_individual}. Among the strongest blazars, they include 3C~279 and NRAO~140 (J0336+3218) in all listed surveys. Furthermore, PKS~1741$-$038 is strong in radio \cite{neutradio2023} and X-rays; it is completely absent from the Swift/BAT survey catalogue but present in the Swift/BAT transient list\footnote{\url{https://swift.gsfc.nasa.gov/results/transients/weak/QSOB1741-038/}}.
\vspace{0.3em}
\item \textsuperscript{3} \url{https://swift.gsfc.nasa.gov/results/transients/weak/QSOB1741-038/}

\end{table*}

\begin{figure}
\centering
\includegraphics[width=0.65\linewidth]{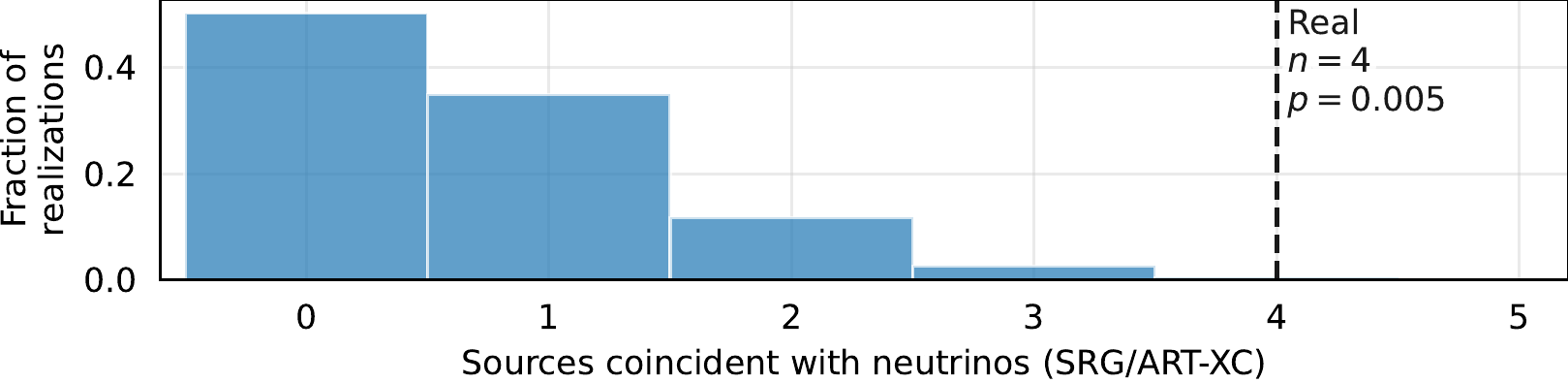}
\vspace{0.2em}

\includegraphics[width=0.65\linewidth]{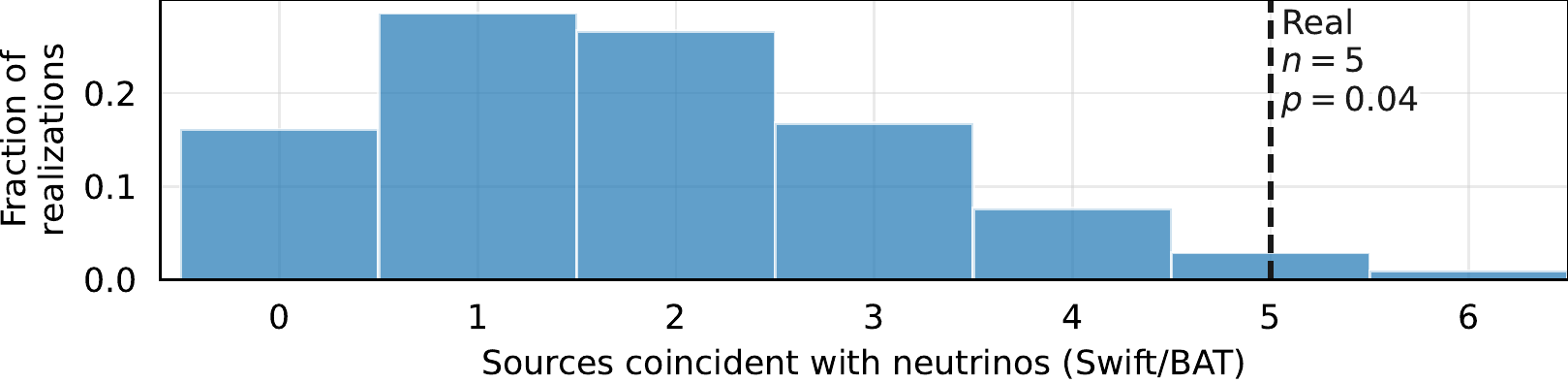}
\vspace{0.2em}

\includegraphics[width=0.65\linewidth]{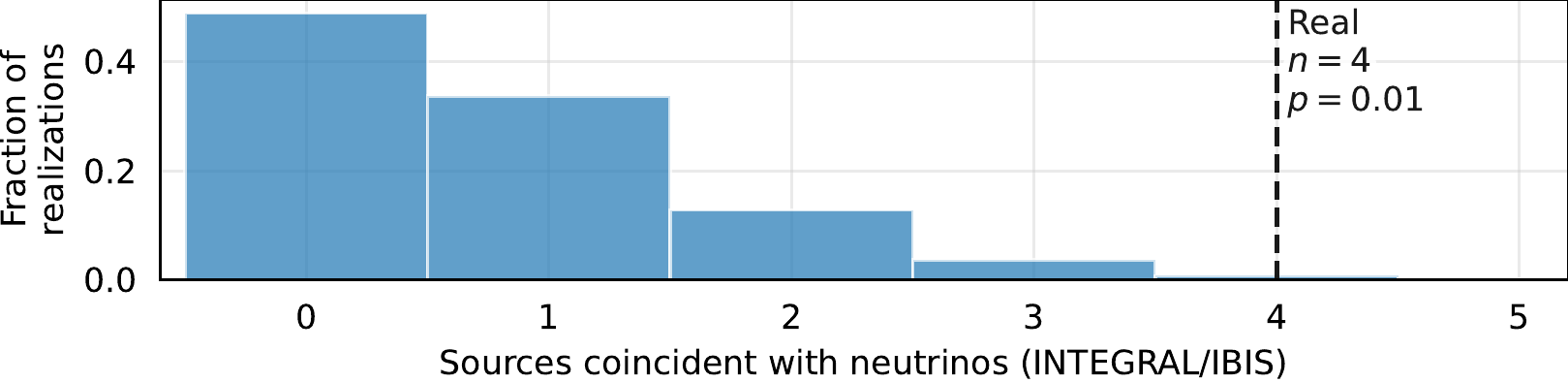}
\vspace{0.2em}

\includegraphics[width=0.65\linewidth]{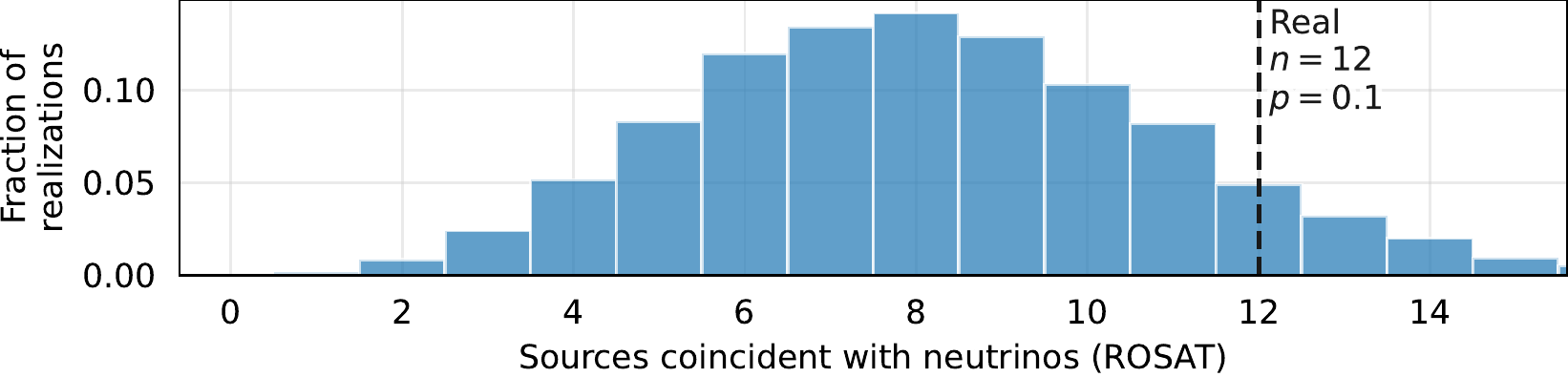}
\caption{
    Statistical analysis visualization: we count matches between neutrinos and X-ray~--~VLBI blazars, and compare the counts to random expectations.
    \label{f:stat_cnts}
}
\end{figure}

\begin{figure*}
\centering
\includegraphics[height=0.3\linewidth,trim=-1mm 0mm -1mm 0mm]{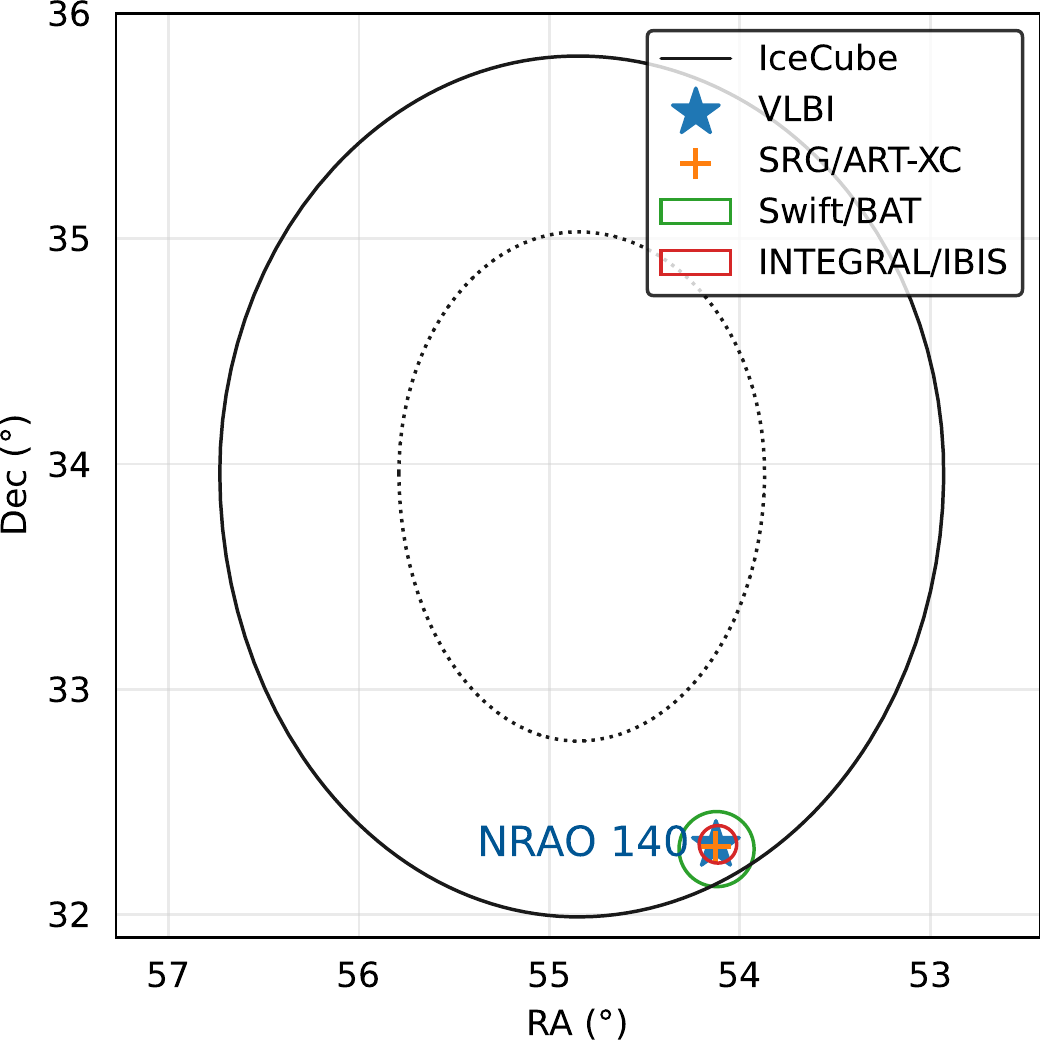}
\includegraphics[height=0.3\linewidth,trim=-1mm 0mm -1mm 0mm]{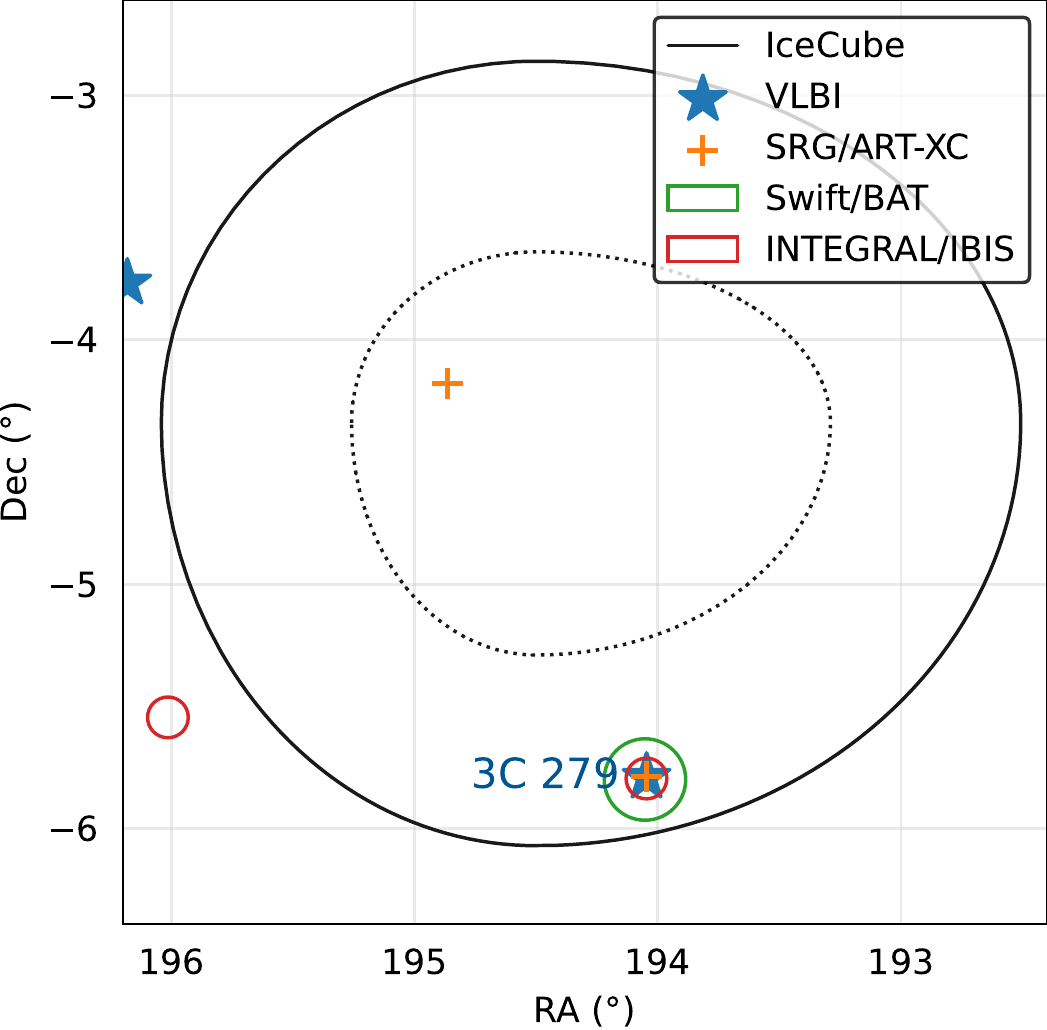}
\includegraphics[height=0.3\linewidth,trim=-1mm 0mm -1mm 0mm]{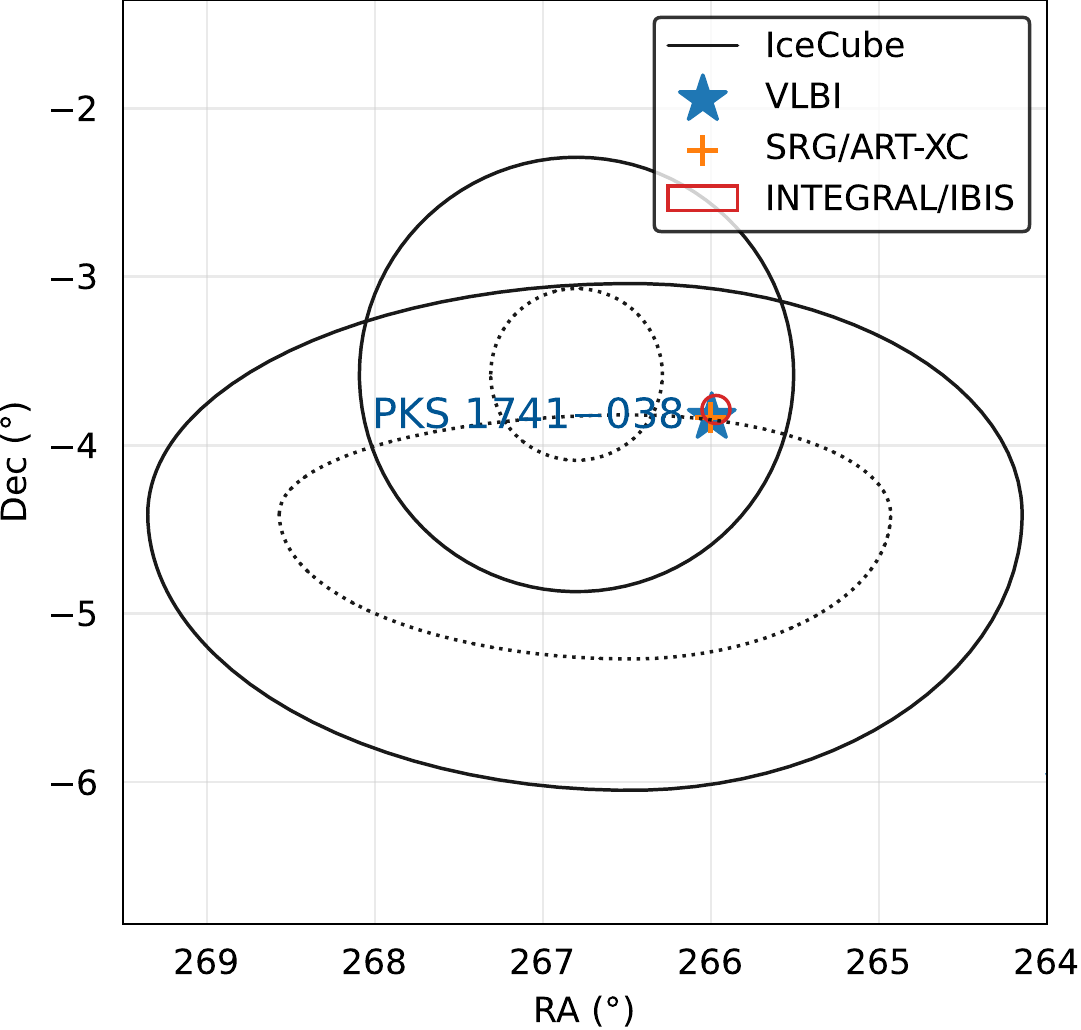}
\caption{
    \label{f:skymaps_individual}
    Sky regions around neutrino events coincident with hard X-ray blazars from the SRG/ART-XC catalog; all three are also present in the INTEGRAL/IBIS sample (\autoref{t:pvalues}).
    The only bright X-ray blazar in each panel is labelled. Error regions are shown for IceCube events, dashed -- original reported uncertainties, solid -- enlarged to account for systematics (\autoref{s:data}). Circles for Swift/BAT and INTEGRAL/IBIS indicate the coincidence regions defined in \autoref{s:Xblazars} and \autoref{t:datasets}. Uncertainties for VLBI and SRG/ART-XC positions are too small to show, their markers sizes don't bear any significance.
}
\end{figure*}

The results of the performed statistical testing procedure~---~counting neutrino-blazar matches~---~are visualized in \autoref{f:stat_cnts}. We obtain a statistical significance of $p=0.5\%$ for the SRG/ART-XC catalog, with 4 neutrino matches compared to $0.7$~expected by chance. These matches consist of two neutrinos from the direction of PKS~1741$-$038 and one each from 3C~279 and NRAO~140. Coincidences for all X-ray catalogs are shown in \autoref{f:skymaps_individual}. They are listed, and statistical outcomes are summarized in \autoref{t:pvalues}. Neither of those $p$-values includes any extra trials, and we do not specifically define the lowest $p$ as our final result: instead, we focus on the ART-XC survey for our statistical analysis a-priori, as described above. Thus, our estimates are not affected by the multiple comparisons issue, and the reported $p$-values do not require any further trial corrections.

All three hard X-ray catalogs~--~SRG/ART-XC, Swift/BAT, and INTEGRAL/IBIS~--~demonstrate some degree of correlation with IceCube neutrinos. We consider the outcomes to be qualitatively consistent between them, given the differences in characteristics such as sky coverage (\autoref{s:data}). Meanwhile, we observe effectively no neutrino correlation at lower-energy X-rays observed by ROSAT. It provides the largest source catalog and all hard X-ray sources are present in the ROSAT sample, but this is compatible with chance coincidences. Limiting the ROSAT catalog to the top~36 sources by their flux for the most direct comparison with SRG/ART-XC (\autoref{t:pvalues}) does not reveal any significant correlation either.

\begin{figure}
\centering
\includegraphics[width=0.6\linewidth]{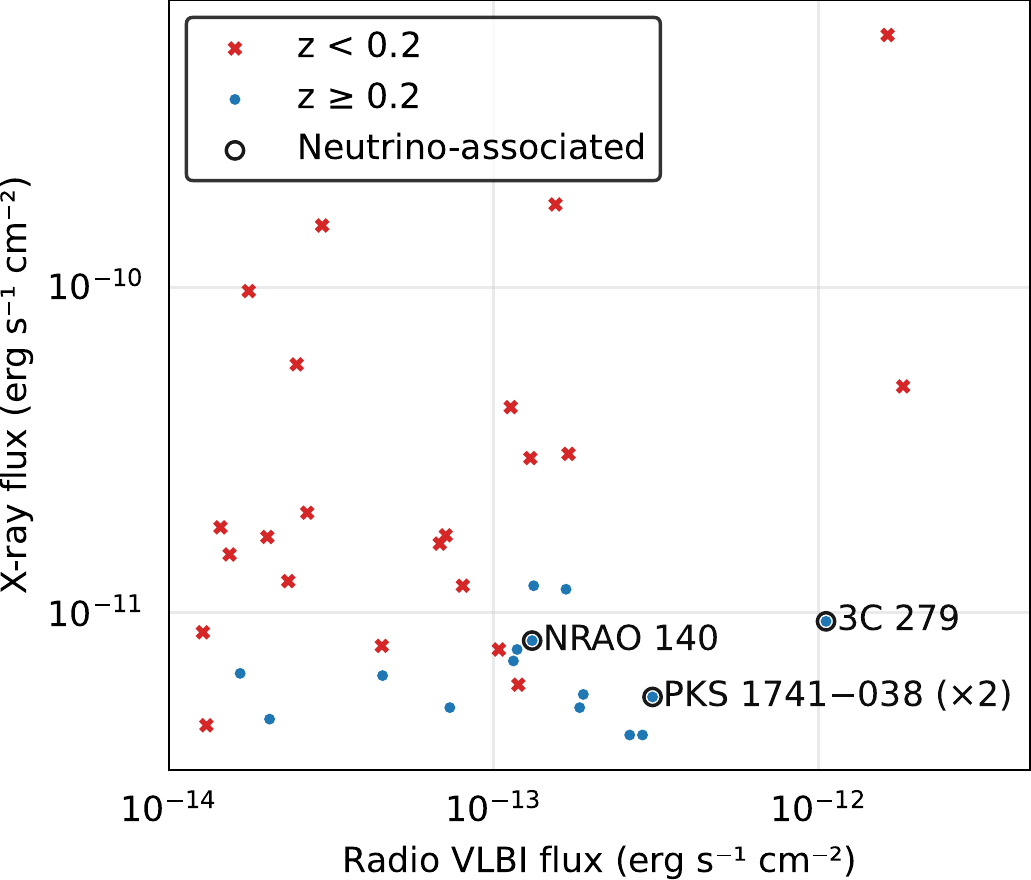}
\caption{
A comparison of flux densities in radio and X-ray bands, for sources in both VLBI and SRG/ART-XC catalogues. The plot demonstrates the lack of strong correlation between the bands, and this observation is consistent for other X-ray catalogues. 
\label{f:fluxcorr}
}
\end{figure}

Based on these directional correlations, we argue that strong X-ray emission at $\sim10$~keV and higher is a characteristic feature of neutrino-emitting blazars. As shown in \autoref{f:fluxcorr}, there is little correlation between the radio and X-ray fluxes of bright blazars. Radio emission and X-rays both correlating with neutrinos is not a manifestation of the same underlying trait, such as blazars being stronger in general.

\subsubsection*{X-ray associations in IceCat-1}
\label{s:icecat}

For the primary analysis presented in \autoref{s:stat}, we utilize the sample of 71 high-energy IceCube events selected following the criteria we originally defined in 2020 (\autoref{s:data_icecube}). This sample was used in a number of works studying neutrino-blazar connection \citep[e.g.,][]{neutradio1,neutradio2023,Hovatta-2021} and demonstrates a correlation steadily growing over time, with more events becoming available.
Recently, IceCube has published the IceCat-1 catalog \cite{2023arXiv230401174A} that uses a different event reconstruction method and differs in how the events were selected. IceCat-1 contains 348 events (340 after applying the IceTop veto) from 2011 to 2023 with estimated energies from 54~TeV to 20~PeV. It is important to evaluate the reliability and consistency of our findings by looking for X-ray blazar associations in the IceCat-1 sample.

We apply the same selection criteria as in \autoref{s:data_icecube} (direction error region $\leq10$~sq.~deg, category of EHE, or estimated $E\geq200$~TeV) and expansion of the reported uncertainties (by~$0.78\degree$) to IceCat events, and obtain 74 matching events. These parameters were motivated and determined in \cite{neutradio1,neutradio2023}. IceCat-1 uses different event reconstruction and energy estimation approaches compared to the catalog utilized in our earlier works and in \autoref{s:stat}, so using the same selection criteria as before could be suboptimal. Still, here we do not perform any additional tuning that would result in unnecessary statistical trials from this comparison.

Exactly as in \autoref{s:stat}, here we look for neutrino associations within the 36 X-ray~--~VLBI matches from the SRG/ART-XC catalog. There are three matches: blazars 3C~279, PKS~1741$-$038, and NRAO~140, yielding a $p$-value of $3\%$.
The other neutrino that we found coincident with the X-ray blazar PKS~1741$-$038 in \autoref{s:stat} is also present in IceCat, but is slightly below our cutoff with its reported energy of 181~TeV.
As noted in \cite{2023arXiv230401174A}, events and associations absent in the IceCat or its subsamples do not contradict their astrophysical nature. The differences indicate the outcome of automatic alert reconstruction procedures.
% This situation is similar to other IceCube reconstruction updates, such as the all-sky point source dataset \cite{icecube10catalog}: as noted in that release, there are no general a-priori reasons to prefer the newer sample over the old one.
% Naturally, results based on earlier data releases are not invalidated, as long as they are performed in a self-consistent manner.

The comparison presented in this section further validates the reliability of associations found in \autoref{s:stat}: the coincident neutrino events are all present in IceCat with similar properties. One of them lies slighly below the energy threshold used in our analysis, with the difference much smaller than typical energy estimation uncertainties. The optimal thresholds to search for astrophysical associations within the IceCat\nobreakdash-1 sample are likely different from what we determined in \cite{neutradio1,neutradio2023}; we leave analysing and determining the differences for future works.

\subsection{The role of blazars in the neutrino -- X-ray association}
\label{s:selection-key}

Earlier works \cite[][and references therein]{neutradio2023} have already demonstrated that high-energy neutrinos are associated with highly-beamed blazars. Our statistical analysis shows that neutrinos are also specifically connected to hard X-ray emission from blazars. To complete the observational picture, it remains to evaluate whether neutrinos can be systematically associated with other bright X-ray sources.

We perform the same statistical testing procedure as described in \autoref{s:stat}, but now utilize X-ray sources that \textit{do not} match with VLBI blazars for direct comparison with \autoref{s:stat} results. We conduct tests for two subsamples of each X-ray catalog:
\begin{itemize}
    \item All non-VLBI-matching sources;
    \item Non-VLBI-matching sources outside of the Galactic plane, $\lvert b\rvert > 10\degree$.
\end{itemize}
We do not find a significant correlation in any of these tests: $p$-values for all X-ray surveys are above 20\%.

The lack of non-blazar X-ray sources correlation with neutrinos, together with the association in \autoref{s:stat}, highlights the importance of identifying blazars when searching for neutrino sources. VLBI observations of parsec-scale radio emission provide a direct observational tracer of jets pointing towards us, that lead to Doppler-boosted electromagnetic emission and potentially similar beaming effects for produced neutrinos \cite[e.g.,][]{Markus23}.

\section{Discussion}
\label{s:disc}

Earlier statistical associations of neutrinos with bright blazars strongly link neutrino production to the very central parsec-scale regions of blazars \cite[e.g.,][]{neutradio1,neutradio2023,Hovatta-2021,2023arXiv230906874A}. Still, observational evidence mostly based on radio emission wasn't enough to say more about the mechanisms of neutrino production. In this work, we demonstrate that blazars associated with neutrinos also have unusually strong hard X-ray emission. As discussed below in this section, this is an important piece of the puzzle and may provide strong constraints for neutrino production modelling.

It is known from previous studies that sources of high-energy neutrinos are numerous, both from general population studies \cite{NeronovSemikozMultiplets,FinleyMultiplets,IceCube-flares} and from estimates of the number of radio blazars contributing to the observed neutrino flux \cite{neutradio2}. Clearly, in this work, we are only able to select a few of the sources, representing the tip of the iceberg. Several probable neutrino sources, such as TXS~0506$+$056, did not show up in our study because they are not among the brightest ones in X-ray (nor in terms of average radio flux, as discussed in \cite{neutradio2023}). We expect that further hard X-ray observations, particularly those continued by SRG/ART-XC, will bring more associations with neutrinos. Here, we do not focus on individual sources but demonstrate statistically that neutrino-associated blazars as a population have higher hard X-ray fluxes than other blazars in the sky.

\subsection{High energy neutrinos from distant blazars}
\label{s:disc:evolution}

\begin{figure}
\centering
\includegraphics[width=0.65\columnwidth,trim=0mm 0mm 4mm 0mm]{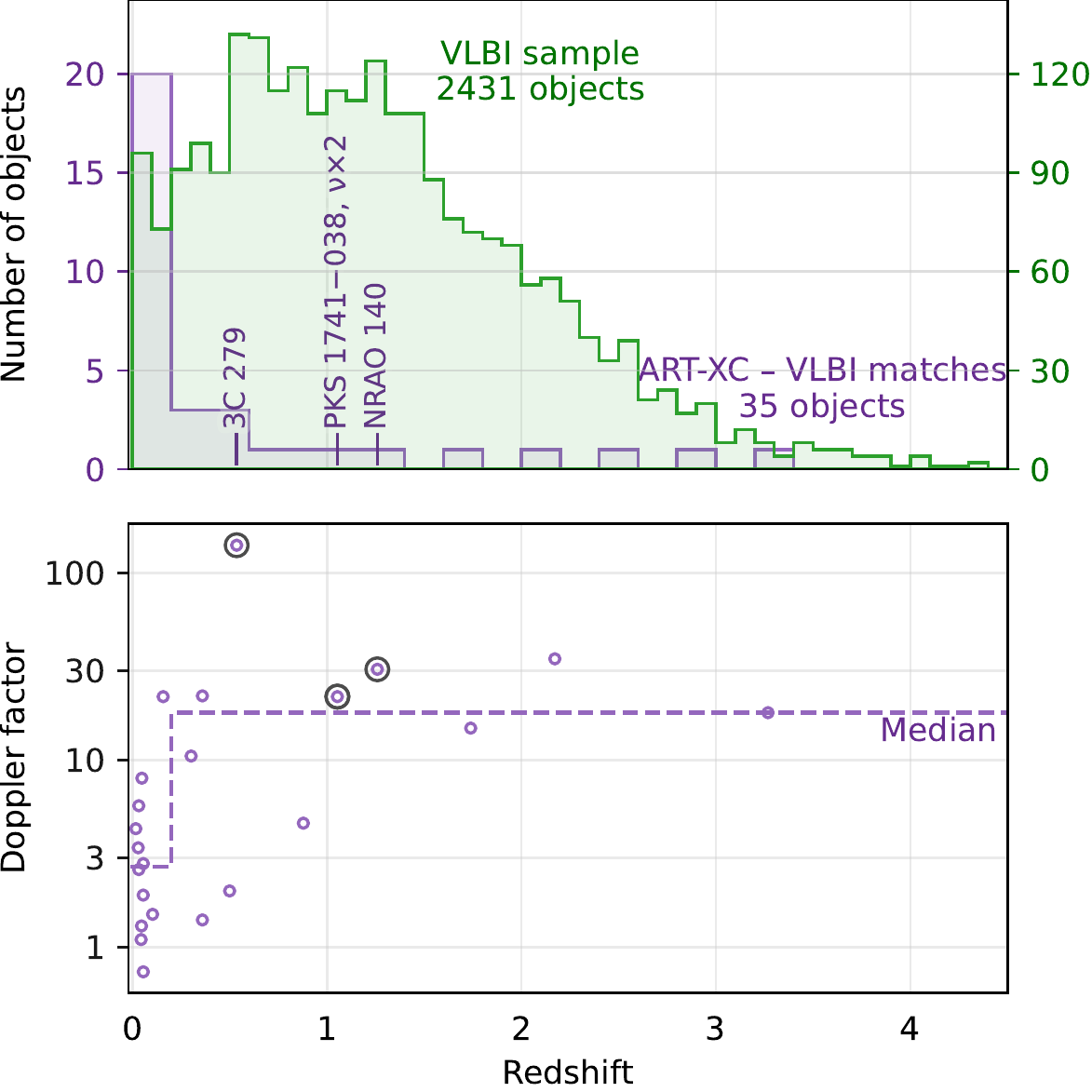}
\caption{
\textit{Top:} The distribution of redshifts for compact sources from the VLBI sample and for X-ray bright sources within it. Note that redshift measurements are only available for a subset of objects from \autoref{t:datasets}. Sources that fall within the error regions of IceCube events are labelled and highlighted. The histogram demonstrates that there are very few X-ray matches beyond $z=0.5$, but all three neutrino coincidences are from those distances.\\
\textit{Bottom:} The Doppler factors of SRG/ART-XC~--~VLBI matches, estimated by \cite{2021ApJ...923...67H}. Purple: all ART-XC -- VLBI matches with measured redshifts and Doppler factors; black: the three neutrino coincidences. Medians are shown separately for nearby sources (first bin in the top panel) and for everything else. Neutrino-coincident sources demonstrate high Doppler boosting even among blazars.
\label{f:redshifts}
}
\end{figure}

Our analysis reveals that all the X-ray blazars associated with neutrinos are located at cosmological distances of $z \sim 1$, as shown in \autoref{f:redshifts} for the SRG/ART-XC sources. This pattern is not a result of any redshift-based preselection, since only a small fraction of the X-ray~--~VLBI matches in the catalog are situated this far from Earth. Observationally, $z \sim 1$ is consistent with earlier works that associate neutrinos with radio-bright blazars \cite[most recently,][]{neutradio2023}.

The observation of radio, X-ray, and neutrino emission from these large distances implies the presence of relativistic beaming effects. Doppler factor measurements by the MOJAVE team \cite{2021ApJ...923...67H} directly probe beaming of the photon emission from the jet; the reliability of those measurements is demonstrated through a remarkable agreement between completely independent approaches. The Doppler factor estimates from \cite{2021ApJ...923...67H} for X-ray~--~VLBI matches are shown in \autoref{f:redshifts}. The jet Doppler factors for the three neutrino-associated blazars are high, $\delta \sim 30\dots100$, even compared to other VLBI blazars. This can indicate that neutrinos undergo similar or stronger beaming effects than the EM emission. See \cite{neutradio2023} for earlier observational hints, \autoref{s:disc:mech} and \cite{Mannheim1,NeronovWhich,Markus23} for more discussion on bulk motion effects on neutrino emission.

Neutrinos can travel across the whole observed Universe, so the flux contribution from sources at different redshifts is defined solely by the density of these sources. Following the cosmological evolution of AGNs, their redshift distribution should be wide, with a maximum at $z \sim 1$; we see this for VLBI blazars directly in \autoref{f:redshifts}.
The bulk of neutrino sources can be expected to reside at these redshifts, as opposed to the local Universe, and our results perfectly agree with such expectations (\autoref{f:redshifts}). Moreover, as discussed in \autoref{s:Xblazars}, our blazar selection approach is not perfect, and at very low redshifts there are disproportionately many AGNs with weak beaming: SRG/ART-XC probably detects non-boosted accretion disk emission there. The observed correlation significance would clearly increase if sources at $z < 0.2$ were dropped, but we do not perform this kind of post-hoc statistical analysis here so as not to introduce multiple trials. The redshift-based selection criteria, $z > 0.2$, is an interesting direction to follow in future neutrino association studies.

\subsection{Multi-messenger emission mechanisms}
\label{s:disc:mech}

\begin{figure}
\centering
\includegraphics[width=0.5\columnwidth]{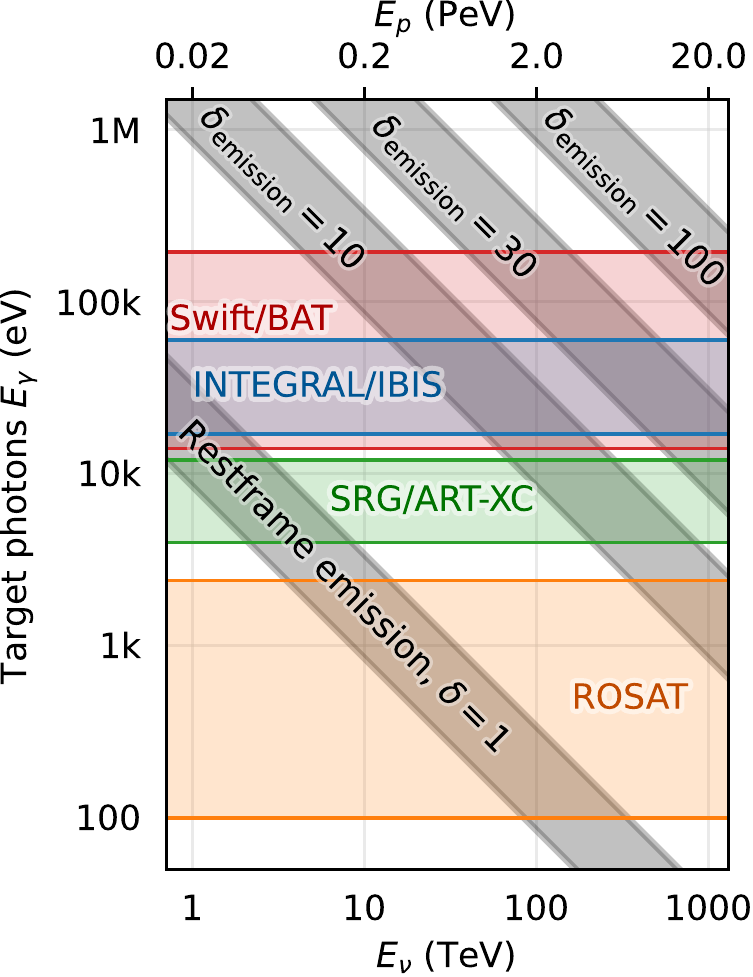}
\caption{
The neutrino energy $E_\nu$ (bottom horizontal axis) versus the energy of relativistic protons $E_p$ (top horizontal axis) and target photons (vertical axis), assuming the neutrino production happens in $p\gamma$ interactions in the $\Delta$-resonance approximation. Both neutrino and photon energies are given in the observer frame. The correspondence between neutrino and target photon energy is represented by diagonal shaded regions for blazar redshifts $0.5 \leq z \leq 2$. The four diagonal bands correspond to the rest-frame photons ($\delta=1$, external radiation), photon emission with Doppler boosting $\delta=10$ typical for radio blazars (\autoref{s:data}), and higher $\delta=30, 100$ that may be more common among neutrino emitters (see \autoref{f:redshifts} and \cite{neutradio2023}). 
The effective energy ranges of X-ray surveys by different observatories are shown as horizontal bands.
\label{f:spectrum}
}
\end{figure}

In this work, we detect the X-ray –- neutrino connection in distant VLBI-selected blazars, whose X-ray emission is dominated by the jet contribution that is strongly Doppler-boosted. These observational results are visualized in \autoref{f:redshifts}, and together with order-of-magnitude expectations from \autoref{f:spectrum} they paint a coherent picture. Indeed, for $\delta_\mathrm{emission} \gtrsim 30$ and $E_\nu \gtrsim 100$~TeV, the target photons in the p$\gamma$ process are expected to arrive at Earth as hard X-rays within the sensitivity ranges of SRG/ART-XC, Swift/BAT, and INTEGRAL/IBIS.

The benchmark relations~(\ref{benchmark_primed_energies}) suggest that hard X-ray photons are a crucial ingredient for efficient neutrino production in blazars. Therefore the results of the present work support both the neutrino-blazar association and the $p\gamma$ neutrino production mechanism often assumed for blazars. To elaborate further on the implications of our observation, recall that the energies of neutrinos and photons in the observer's frame are given by
\begin{equation}
E_{\gamma,\nu}=\frac{\delta}{1+z} E'_{\gamma,\nu},
\end{equation}
where $z$ is the source's redshift and $\delta$ is the Doppler factor of the frame where target photons are isotropic; see \cite{Markus23} for a more detailed discussion. Using Eq.~(\ref{benchmark_primed_energies}), we find  that the observed energies of the target photons are
\begin{equation}
E_\gamma \simeq \mbox{8.5~keV}\,
\left(\frac{\delta}{10}\right)^{2}
\left(\frac{2}{1+z}\right)^{2}
\left(\frac{E_\nu}{200~\mathrm{TeV}}\right)^{-1}.
\end{equation}
This relation is visualized in \autoref{f:spectrum} for a range of $\delta$ and $z$ values. For a typical neutrino-associated blazar with $z\sim 1$ and $\delta \gtrsim 10$ (\autoref{f:redshifts}), the target photon energies fall nicely in the SRG/ART-XC sensitivity band and above, provided the target photons are produced in the jet. On the contrary, if the target photons are emitted in the rest frame of the central black hole (e.g., from the disk corona), their energies would be two orders of magnitude lower. In addition, in the latter case, the flux of target photons would not be Doppler-boosted and likely not detected by current telescopes.

These estimates indicate that the observations presented in this work support the $p\gamma$ neutrino production mechanism in blazars. Furthermore, they form an argument that the origin of the target photons is in the Doppler-boosted regions close to the jet base. This explanation fits well with models where the target photons constitute the synchrotron-self-Compton (SSC) emission in the jet up to parsec scales \cite{neutradio2,Polina:RadioModel}. This scenario also provides a natural explanation for the parsec-scale radio flux density correlation with the neutrino production rate, since the amount of SSC photons, and hence the neutrino flux, is proportional to the amount of synchrotron photons that emit radio waves. Models in which the target photons are produced in the accretion disk or its hot corona can therefore be less favored for the production of neutrinos above 200~TeV, as studied here. Still, our results alone do not conclusively rule out these or other potential scenarios and mechanisms. Evaluating the potential connection between X-ray flares and neutrino emission may also shed more light and allow distinguishing between alternative models: see, e.g., \cite{2022MNRAS.510.4063S}.

At higher photon energies, GeV gamma rays, no systematic correlation of blazar emission with neutrinos was found, see e.g. \cite{2018A&A...620A.174K,2019ICRC...36..916H,2023ApJ...954...75A}. Gamma-ray photons fundamentally accompany neutrinos on production, but they efficiently interact with low-energy photons ($\lesssim 1$~keV) producing electron-positron pairs and cascading down in energy. Cascading remains efficient until about a GeV, see discussion and references in \cite{neutradio2}. Therefore, regions with efficient neutrino production are likely opaque to gamma ray emission and we cannot observe it directly. X-ray observations potentially provide a more direct tracer of neutrino production.

\section{Summary}
\label{s:summary}

In this paper, we observationally and systematically evaluate the connection between X-ray emission from blazars and high-energy neutrinos detected by IceCube. We utilize all-sky catalogs in the radio band produced by VLBI and in the X-ray band, with special care taken to make the samples as uniform as possible. X-ray catalogs were produced by SRG/ART-XC, Swift/BAT, and INTEGRAL/IBIS observations in hard X-rays, as well as by ROSAT at lower energies.

We find a correlation between hard X-ray blazar emission at around $E \sim 10$~keV and IceCube high-energy neutrinos. For the SRG/ART-XC catalog, the chance probability of the correlation with IceCube events selected in \autoref{s:data_icecube} is 0.5\% (see \autoref{t:pvalues}); this result is validated using the IceCat-1 sample, resulting in the chance probability of 3\% (\autoref{s:icecat}). Meanwhile, softer X-rays at $E \lesssim 1$~keV, do not demonstrate any neutrino correlation (\autoref{t:pvalues}). Neutrinos are more likely to arrive from highly beamed blazars with strong X-ray emission, indicating a physical association, possibly a common origin mechanism.

These results are of a great help in narrowing down neutrino production scenarios. Assuming the neutrinos are born in proton-photon interactions, the observed X-rays could come from the population of target photons in these processes. The bulk of X-rays from blazars at cosmological distances comes from their beamed jet emission. If the observed X-rays are indeed physically related to neutrino production, this is a potential argument for target photons being produced in the jet, for instance through self-Compton scattering. Still, scenarios with external target photons might be more energetically favorable, and this tension is a prime question for further studies.

We emphasize the importance of selecting Doppler-boosted blazars for the purpose of neutrino associations. Parsec-scale radio emission measured by VLBI is a reliable tracer of relativistic beaming in the jet, and beaming also affects the produced neutrinos. Potentially, observing hard X-rays can be a way to probe neutrino production processes directly, while VLBI selects highly-beamed blazar jets so that neutrinos are more likely to be detected on Earth.

We anticipate that the current and future growth of neutrino observatories will bring about more neutrino detections, and their parameters will be more accurately measured. Electromagnetic observatories also continue observing; specifically, the focus of SRG/ART-XC was on the all-sky survey, and the updated catalogs are expected to become deeper and more precise. Closely monitoring neutrino-associated blazars, and evaluating the potential connection between X-ray flares and neutrino detections, would be of a great interest as well. Together, these advancements would allow for a deeper quantitative evaluation of the neutrino-blazar associations, and further narrow down neutrino production mechanisms.

\acknowledgments
We thank Sergey Sazonov for helpful comments and discussions on various parts of this work, as well as Eduardo Ros for comments on the manuscript.
AVP was supported by the Black Hole Initiative, which is funded by grants from the John Templeton Foundation (Grant~\#60477, 61479, 62286) and the Gordon and Betty Moore Foundation (Grant~GBMF\nobreakdash-8273).
RAB and AAL acknowledge the support by the RFBR grant 19-29-11029. AAS was partly supported by the project number 0033-2019-0005 of the Russian Ministry of Science and Higher Education.
YYK was supported by the M2FINDERS project, which has received funding from the European Research Council (ERC) under the European Union’s Horizon 2020 Research and Innovation Programme (grant agreement No~101018682).

\bibliographystyle{JHEP}
\bibliography{nuxray}{}

\end{document}